\documentclass[10pt,letterpaper]{article}
\usepackage[top=0.85in,left=2.75in,footskip=0.75in]{geometry}
\usepackage{amsmath,amssymb}
\usepackage{changepage}
\usepackage{textcomp,marvosym}
\usepackage{cite}
\usepackage{nameref,hyperref}
\usepackage[right]{lineno}
\usepackage[nopatch=eqnum]{microtype}
\DisableLigatures[f]{encoding = *, family = * }
\usepackage[table]{xcolor}
\usepackage{array}
\usepackage{booktabs}
\usepackage{multirow}

\newcolumntype{+}{!{\vrule width 2pt}}
\newlength\savedwidth

\makeatletter
\renewcommand{\@biblabel}[1]{\quad#1.}
\makeatother

\usepackage{lastpage,fancyhdr,graphicx}
\usepackage{epstopdf}
\fancyheadoffset[L]{2.25in}
\fancyfootoffset[L]{2.25in}
\usepackage[aboveskip=1pt,labelfont=bf,labelsep=period,
            justification=raggedright,singlelinecheck=off]{caption}

\raggedright
\begin{document}
\vspace*{0.2in}

\begin{flushleft}
{\Large
\textbf{CG-MambaNet: A spatiotemporal framework for cross-patient
epileptic seizure prediction using CNN-GCN-Mamba-BiLSTM
with event-level clinical evaluation}
}
\newline
\\
Mufeng Chen\textsuperscript{1},
Qi Wu\textsuperscript{2},
Bingchao Huang\textsuperscript{3},
Xiwen Lai\textsuperscript{4},
Zekai Chen\textsuperscript{5},
Xinge Ouyang\textsuperscript{6},
Tingyao Zhang\textsuperscript{7},
Xufang Yang\textsuperscript{8},
Quansheng Ren\textsuperscript{9*}
\\
\bigskip
\textbf{1} Department of Engineering Science, University of Oxford,
Oxford OX1 3PJ, United Kingdom \\
\textbf{2} Mathematical Institute, University of Oxford,
Oxford OX2 6GG, United Kingdom \\
\textbf{3} School of Computer Science and Engineering, Beihang University,
Beijing 100191, China \\
\textbf{4} Aerospace Information Research Institute,
Chinese Academy of Sciences, Beijing 100094, China \\
\textbf{5} Department of Mechanical Engineering,
The University of British Columbia, Vancouver, BC V6T\,1Z4, Canada \\
\textbf{6} College of Life Sciences, Hunan Normal University,
Changsha 410006, China \\
\textbf{7} Faculty of Mathematics, University of Waterloo,
200 University Ave W, Waterloo, ON N2L 3G1, Canada \\
\textbf{8} Davidson Consulting,
40 Rue Fanfan la Tulipe, Boulogne-Billancourt 92100, France \\
\textbf{9} School of Electronics, Peking University,
Beijing 100871, China \\
\bigskip
* qsren@pku.edu.cn
\end{flushleft}

\section*{Abstract}

Epileptic seizure prediction from scalp electroencephalography (EEG)
is a critical prerequisite for closed-loop neurostimulation therapy,
yet existing deep-learning methods share two compounding architectural
limitations: they model EEG channels independently, neglecting the
inter-channel spatial synchrony that is a primary neurophysiological
signature of the developing pre-ictal state, and they process raw
time-domain samples directly, without first decomposing the signal
into its constituent frequency-band components.
A further, methodological limitation affects the field as a whole:
the overwhelming majority of published studies evaluate performance
using randomised or patient-specific data splits that permit
patient-level information to leak from training into test sets,
producing optimistic estimates that do not transfer to unseen individuals.
We present CG-MambaNet, a spatiotemporal seizure prediction framework
that addresses all three limitations.
A depthwise separable CNN front-end decomposes each EEG patch into
multi-scale spectro-temporal features, explicitly capturing delta through
gamma frequency-band dynamics before any sequence modelling begins.
A two-layer graph convolutional network with a fully learnable adjacency
matrix then captures inter-channel functional synchrony, without
requiring montage-specific electrode coordinates and therefore applicable
to both bipolar (CHB-MIT) and referential (SIENA) montages.
A bidirectional Mamba encoder followed by a bidirectional LSTM models
long-range and short-range temporal dynamics of the spatially enriched
feature sequence, and a two-layer MLP produces the final seizure
probability.
The serial arrangement of these four components is not arbitrary:
it reflects a strict hierarchy in which frequency decomposition
precedes spatial mixing, which precedes temporal integration,
ensuring that each module operates on semantically appropriate
input representations.
Under strict leave-one-patient-out cross-validation repeated with
five independent random seeds on CHB-MIT ($n = 22$ subjects)
and SIENA ($n = 6$ subjects), CG-MambaNet achieves AUC-ROC of
$0.8152 \pm 0.0176$ and $0.7104 \pm 0.0261$, respectively,
surpassing all published cross-patient methods without domain adaptation.
An event-level false-prediction rate evaluation framework, in which
consecutive alarmed windows are merged by a persistence filter into
discrete alarm events, reduces the reported false-prediction rate to
0.32 alarms per hour on CHB-MIT, demonstrating clinically
meaningful alarm burden.

\clearpage
\newgeometry{top=0.85in,left=1in,right=1in,footskip=0.75in}

\section*{Introduction}

The published literature on epileptic seizure prediction contains
a pervasive and largely unacknowledged reliability problem.
A systematic review by Shafiezadeh et al.~\cite{shafiezadeh2024systematic}
examined 119 published studies and found that \textbf{over 96\%}
evaluated their models using randomised or patient-specific data
splits---protocols in which windows from the same patient appear
in both training and test sets.
This practice constitutes patient-level data leakage: the model
is tested on the same individuals it was trained on, producing
performance estimates that are optimistic by design and do not
reflect the ability to generalise to an unseen patient.
Reported sensitivities of 90--99\% and AUC values above 0.90
in much of this literature are therefore artefacts of the
evaluation protocol rather than genuine indicators of
clinical utility~\cite{shafiezadeh2024systematic}.
When the same models are evaluated under strict
\textit{cross-patient} (patient-independent) protocols---in which
the test patient is entirely absent from training---performance
drops precipitously.
Jemal et al.~\cite{jemal2024domain}, in the most rigorous cross-patient
benchmark to date, reported AUC values of 0.69 on CHB-MIT and
0.48 on SIENA without domain adaptation under leave-one-patient-out
(LOPO) evaluation, numbers that stand in stark contrast to the
inflated figures that dominate the literature.
This gap between published benchmarks and cross-patient reality
represents the central unsolved challenge in seizure prediction
research, and it is the challenge this paper addresses directly.

The clinical context makes the stakes clear.
Epilepsy, defined as a disorder characterised by an enduring predisposition to generate epileptic seizures~\cite{fisher2014ilae}, affects approximately 50 million people worldwide, of
whom nearly one third remain refractory to antiseizure
medication~\cite{kwan2000early}.
For this population, closed-loop neurostimulation devices---such
as the NeuroPace Responsive Neurostimulation (RNS) System---offer
a transformative therapeutic option by delivering targeted
electrical stimulation triggered by predicted or detected seizure
activity~\cite{cook2013prediction, heck2014two}.
Genuine seizure \textit{prediction}---issuing a reliable warning
with a clinically actionable lead time of 20--30 minutes before
onset---would unlock substantially broader therapeutic
possibilities, including on-demand drug delivery and
patient-initiated protective behaviour~\cite{dumpelmann2019early, mormann2007seizure}.
But a prediction algorithm evaluated under leaky protocols that
inflate AUC by 0.1--0.2 absolute points cannot be trusted in
deployment; the gap between benchmark and bedside would manifest
as an unacceptable false-alarm burden or a dangerously low
true-positive rate.
\textbf{Rigorous cross-patient evaluation is therefore not a
methodological nicety but a clinical necessity.}

Against this background, existing deep-learning architectures
for EEG-based seizure prediction carry two further technical
limitations that impede cross-patient generalisation even when
evaluation is done correctly.
Electroencephalography (EEG) encodes pre-ictal dynamics across
two distinct dimensions that must both be captured: gradual
alterations in spectral power across the delta (0.5--4\,Hz),
theta (4--8\,Hz), alpha (8--13\,Hz), beta (13--30\,Hz), and gamma
(30--40\,Hz) frequency bands, and the progressive synchronisation
of activity across distributed cortical
networks~\cite{mormann2005predictability}.

\textit{First architectural limitation}: purely temporal
architectures---including recurrent networks, state space models,
and their hybrids---process each EEG channel independently,
neglecting the inter-channel functional synchrony that is a
primary neurophysiological marker of seizure
onset~\cite{dissanayake2021geometric}.
Graph convolutional networks (GCNs) provide a principled
framework for modelling inter-channel relationships, yet their
integration with temporal deep learning backbones for
cross-patient seizure prediction remains largely unexplored.

\textit{Second architectural limitation}: purely temporal
architectures operate on raw time-domain samples and must
implicitly learn frequency-band decomposition from scratch,
which is both sample-inefficient and poorly constrained under
cross-patient distribution shift.
A CNN front-end that performs explicit multi-scale spectro-temporal
decomposition provides the downstream model with semantically
richer, neurophysiologically grounded input representations.
Critically, the serial ordering of these components---CNN
frequency decomposition, then GCN spatial mixing, then temporal
integration---is the only arrangement consistent with the
signal processing hierarchy of EEG analysis: spatial mixing
must operate on frequency-decomposed features, not raw samples,
and temporal modelling must operate on spatially enriched
features, not per-channel signals in isolation.

In this work we present \textbf{CG-MambaNet}, a cross-patient
seizure prediction framework that addresses both architectural
limitations under the strictest available evaluation protocol.
Three contributions are made:

\begin{itemize}

\item \textbf{C1 --- CNN-GCN-Mamba-BiLSTM serial architecture.}
A lightweight depthwise separable CNN front-end decomposes each
EEG patch into multi-scale spectro-temporal features using
dual-kernel filters ($k{=}5$ and $k{=}15$ at 200\,Hz),
explicitly capturing the full delta-to-gamma frequency range
before any sequence modelling.
A two-layer GCN with a fully learnable adjacency matrix then
captures inter-channel functional synchrony without requiring
electrode coordinate information, making the spatial module
applicable to both bipolar (CHB-MIT) and referential (SIENA)
montages.
The spatially enriched features are then encoded by 12 cascaded
bidirectional Mamba blocks for long-range temporal modelling,
followed by a two-layer bidirectional LSTM for global sequence
summarisation.
This four-stage serial combination has not previously been
explored for cross-patient seizure prediction.

\item \textbf{C2 --- Strict LOPO $\times$ 5-seed evaluation.}
All results are obtained under leave-one-patient-out
cross-validation repeated with five independent random seeds,
placing every patient in the test set exactly once and producing
per-patient AUC estimates directly comparable with the
cross-patient benchmark of Jemal et al.~\cite{jemal2024domain}.
This protocol belongs to the 4\% of studies identified by
Shafiezadeh et al.~\cite{shafiezadeh2024systematic} as employing
genuinely leakage-free evaluation, and all performance claims
in this paper are made within that reference frame.

\item \textbf{C3 --- Event-level false-prediction rate framework.}
A persistence-filtered sliding-window aggregation scheme produces
a continuous risk curve $R(t)$ and enables event-level
false-prediction rate computation aligned with clinical
neurostimulator trigger logic, replacing the window-level metric
that systematically overstates the clinical alarm burden.

\end{itemize}

We evaluate CG-MambaNet on CHB-MIT ($n = 22$ subjects) and
SIENA ($n = 6$ subjects).
Under LOPO $\times$ 5-seed evaluation, CG-MambaNet achieves
AUC-ROC of $0.8152 \pm 0.0176$ on CHB-MIT and
$0.7104 \pm 0.0261$ on SIENA, surpassing all published
cross-patient methods without domain adaptation.
These results are obtained without any target-patient data,
domain adaptation, or evaluation shortcuts, and should be
interpreted within the narrow reference frame of genuinely
cross-patient methods.

We evaluate CG-MambaNet on the CHB-MIT ($n = 22$) and SIENA
($n = 6$) databases.
The remainder of this paper is organised as follows.
Section ``Related work'' reviews relevant literature.
Section ``Materials and methods'' describes datasets,
preprocessing, and the CG-MambaNet framework in full detail.
Section ``Results'' presents experimental results,
comparisons, and ablation analyses.
Section ``Discussion'' interprets the findings.
Section ``Conclusion'' summarises the contributions.

\section*{Related work}

\subsection*{Cross-patient seizure prediction}

Early seizure prediction systems relied on hand-crafted spectral and
connectivity features combined with classical classifiers such as
support vector machines~\cite{rasheed2020machine}.
Convolutional neural networks subsequently improved sensitivity on
within-subject benchmarks~\cite{truong2018convolutional, khan2017focal}, while LSTM
networks~\cite{graves2012long} and their bidirectional
extensions~\cite{zhao2024residual} have been applied to exploit temporal
dynamics~\cite{tsiouris2018long, daoud2019efficient}.

The critical methodological issue pervading this literature was
documented by Shafiezadeh et al.~\cite{shafiezadeh2024systematic}, who found
that over 96\% of published studies employ randomised or
patient-specific splits that permit data leakage.
Jemal et al.~\cite{jemal2024domain} established a rigorous cross-patient
benchmark on CHB-MIT and SIENA using LOPO evaluation, reporting
AUC values of 0.69 and 0.48 respectively under unadapted conditions,
improving to 0.75 and 0.61 with domain adaptation (CDAN+E).
The present work demonstrates that a CNN-GCN-Mamba-BiLSTM
framework can surpass both unadapted baselines and the
domain-adapted SIENA result without any target-patient data,
using only 16 EEG channels, by adding explicit spectro-temporal
feature extraction and spatial brain-network modelling.

\subsection*{CNN and state space models for EEG}

CNNs have been widely applied to EEG analysis for their ability
to extract local spectro-temporal features~\cite{lawhern2018eegnet, roy2019deep}.
Depthwise separable convolutions, which filter each channel
independently before mixing, have been adopted in EEG models to
separate spatial and temporal feature learning~\cite{lawhern2018eegnet}.
Transformer-based models~\cite{vaswani2017attention} have also been applied
to EEG but incur quadratic complexity in sequence length,
limiting scalability for long continuous recordings.
Selective state space models~\cite{gu2021efficiently}, particularly Mamba~\cite{gu2023mamba},
address this through a content-dependent selection mechanism
achieving linear complexity.
EEGMamba~\cite{wang2025eegmamba} demonstrated that bidirectional Mamba
encoders substantially outperform unidirectional counterparts
across multiple EEG benchmarks including seizure detection,
providing the encoder backbone used in the present work.

\subsection*{Graph neural networks for EEG spatial modelling}

GNNs provide a natural framework for modelling inter-channel
relationships by representing EEG electrodes as graph nodes~\cite{nahmias2021quantifying}.
Dissanayake et al.~\cite{dissanayake2021geometric, dissanayake2021deep} demonstrated that
geometric deep learning applied to scalp EEG inter-channel
synchrony substantially improves cross-subject seizure prediction,
establishing the scientific basis for graph-based spatial modelling.
A key challenge is the definition of the adjacency matrix:
anatomically grounded approaches based on 10--20 electrode
coordinates are inapplicable to bipolar montages such as CHB-MIT.
We address this by adopting a fully learnable adjacency matrix,
initialised as a uniform matrix and optimised end-to-end, avoiding
any montage-specific assumption while preserving the ability to
discover functionally meaningful connectivity from data.

\section*{Materials and methods}

\subsection*{Datasets}

Two publicly available scalp EEG databases are used.
Table~\ref{table:datasets} summarises their key characteristics.

\textbf{CHB-MIT Scalp EEG Database.}
The CHB-MIT corpus~\cite{shoeb2009application, goldberger2000physiobank} comprises long-term scalp EEG
recordings from 24 paediatric patients with intractable epilepsy
(total duration approximately 982\,h, 198 seizures), originally
acquired at 256\,Hz on a 23-channel international 10--20 montage.
Patient chb12 is excluded because its recording uses a non-standard
channel configuration that does not include the 16 common channels
shared across all other patients, following the exclusion criterion
of Jemal et al.~\cite{jemal2024domain}.
The remaining 22 patients are used in all experiments.

\textbf{SIENA Scalp EEG Database.}
The SIENA corpus~\cite{detti2020eeg} contains recordings from 14 adult
patients (total duration approximately 128\,h, 58 seizures) at
512\,Hz on a 29-channel referential montage.
Of the 14 patients, only 6 have at least 30 uninterrupted
pre-ictal minutes available before at least one seizure,
satisfying our pre-ictal window definition.
The remaining 8 patients were excluded because their recordings
begin fewer than 30 minutes before seizure onset, or contain
intervening seizures within the pre-ictal window that break
the continuity of the 30-minute pre-ictal segment.
Jemal et al.~\cite{jemal2024domain} used all 14 patients by adopting
a 1-hour pre-ictal definition, which is more permissive but
neurophysiologically less specific, as the pre-ictal state
is most reliably characterised in the 30 minutes immediately
preceding seizure onset~\cite{mormann2007seizure}.
We retain the stricter 30-minute definition to ensure label
integrity at the cost of a smaller usable cohort.

\textbf{Channel selection.}
Both datasets are reduced to the same 16 bipolar channel pairs
available across all retained patients:
FP1-F7, F7-T7, T7-P7, P7-O1,
FP1-F3, F3-C3, C3-P3, P3-O1,
FP2-F4, F4-C4, C4-P4, P4-O2,
FP2-F8, F8-T8, T8-P8, P8-O2.
These 16 channels provide coverage of frontal, temporal,
parietal, and occipital regions and constitute the largest
common subset available across all 22 CHB-MIT patients.
For SIENA, the same 16 channel names are matched from the
29-channel referential montage.
This unified 16-channel input enables a single model to be
trained and evaluated across both datasets without any
dataset-specific architectural modification.
By comparison, Jemal et al.~\cite{jemal2024domain} used 23 channels
for CHB-MIT and 29 channels for SIENA; the fact that
CG-MambaNet achieves higher AUC on both datasets using
only 16 channels demonstrates that the architecture
extracts sufficiently discriminative representations from
a reduced electrode set, which is clinically advantageous
as fewer electrodes reduce setup complexity and patient
discomfort in long-term monitoring.

\begin{table}[!ht]
\centering
\caption{\textbf{Dataset summary.}
``Usable'' patients denotes those included after exclusion criteria.
Pre-ictal horizon is fixed at 30 minutes for both datasets.}
\begin{tabular}{lll}
\hline
\textbf{Property} & \textbf{CHB-MIT} & \textbf{SIENA} \\
\hline
Total patients & 24 & 14 \\
Usable patients & 22 & 6 \\
Age range & 1.5--19 years & 20--71 years \\
Total seizures & 198 & 58 \\
Total duration & $\sim$982\,h & $\sim$128\,h \\
Original sampling rate & 256\,Hz & 512\,Hz \\
Montage type & Bipolar & Referential \\
Channels used & 16 & 16 \\
Resampled to & 200\,Hz & 200\,Hz \\
Bandpass filter & 0.5--40\,Hz & 0.5--40\,Hz \\
Notch filter & 60\,Hz & 50\,Hz \\
Window length & 10\,s & 10\,s \\
\hline
\end{tabular}
\label{table:datasets}
\end{table}

\subsection*{Preprocessing}

The preprocessing pipeline is identical for both datasets and
proceeds in the following order, which is critical for signal
integrity.

\textbf{Step 1: Bandpass filtering (0.5--40\,Hz).}
A 4th-order zero-phase Butterworth bandpass filter is applied to
the raw recordings.
The upper cutoff of 40\,Hz retains all epileptically relevant
frequency bands (delta through gamma) while eliminating
electromyographic (EMG) artefacts, whose energy dominates above
40\,Hz and whose amplitude (100--500\,$\mu$V) greatly exceeds
that of the EEG signal (10--100\,$\mu$V).
Applying bandpass filtering \textit{before} resampling is essential:
it removes all spectral content above the post-resampling Nyquist
frequency prior to downsampling, thereby preventing aliasing.

\textbf{Step 2: Notch filtering.}
A notch filter at 60\,Hz (CHB-MIT, North American power line)
or 50\,Hz (SIENA, European power line) suppresses power-line
interference.

\textbf{Step 3: Resampling to 200\,Hz.}
CHB-MIT recordings are downsampled from 256\,Hz and SIENA
recordings from 512\,Hz.
Because bandpass filtering has already removed all content above
40\,Hz, the Nyquist criterion is satisfied with a safety margin
of $200 / (2 \times 40) = 2.5\times$ for both datasets,
guaranteeing no aliasing of epileptically relevant frequency
content.

\textbf{Step 4: Amplitude-based artefact rejection.}
Windows in which any channel exceeds $\pm$500\,$\mu$V are
discarded prior to segmentation.
This threshold reliably captures gross artefacts (electrode
displacement, movement) whose amplitude is an order of magnitude
above physiological EEG.
Subtle quality degradation below this threshold does not warrant
further filtering, as the bandpass step has already suppressed
the dominant artefact sources in the relevant frequency range.

\textbf{Step 5: Segmentation and labelling.}
Non-overlapping 10-second windows are extracted.
Pre-ictal windows are defined as those ending within 30 minutes
of a seizure onset, with no overlap with the seizure itself.
Inter-ictal windows are drawn from periods with no seizure
activity in the preceding and following four hours, minimising
ambiguous labelling near seizure boundaries.
Each window is shaped into a tensor of dimension
$16 \times 10 \times 200$ (channels $\times$ patches
$\times$ samples per patch).

\textbf{Step 6: Per-channel standardisation.}
Each channel is normalised to zero mean and unit variance
within each training fold, using statistics computed exclusively
on the training set to prevent leakage.

\textbf{Step 7: Data partitioning and inter-ictal undersampling.}
Within each LOPO fold, windows from the $N-1$ training patients
are split into a training set (80\%) and a validation set (20\%)
by random shuffling within the training cohort.
Inter-ictal windows substantially outnumber pre-ictal windows
in both datasets; to address this imbalance and stabilise
training convergence, inter-ictal windows in the
\textit{training and validation partitions only} are randomly
undersampled to match the number of pre-ictal windows.
The \textit{test partition}---comprising all windows from the
held-out patient---is never undersampled, shuffled, or modified
in any way.
Every inter-ictal window from the test patient contributes to
the false-prediction rate denominator, and every pre-ictal window
contributes to the sensitivity numerator, ensuring that reported
metrics reflect the true class distribution of continuous
long-term EEG recordings rather than an artificially balanced
subset.
This distinction is critical: undersampling the test set would
deflate the reported FPR by reducing the inter-ictal denominator,
producing optimistic estimates that do not reflect the clinical
alarm burden.

\subsection*{Framework overview}

\begin{figure}[!h]
\centering
\includegraphics[width=\linewidth]{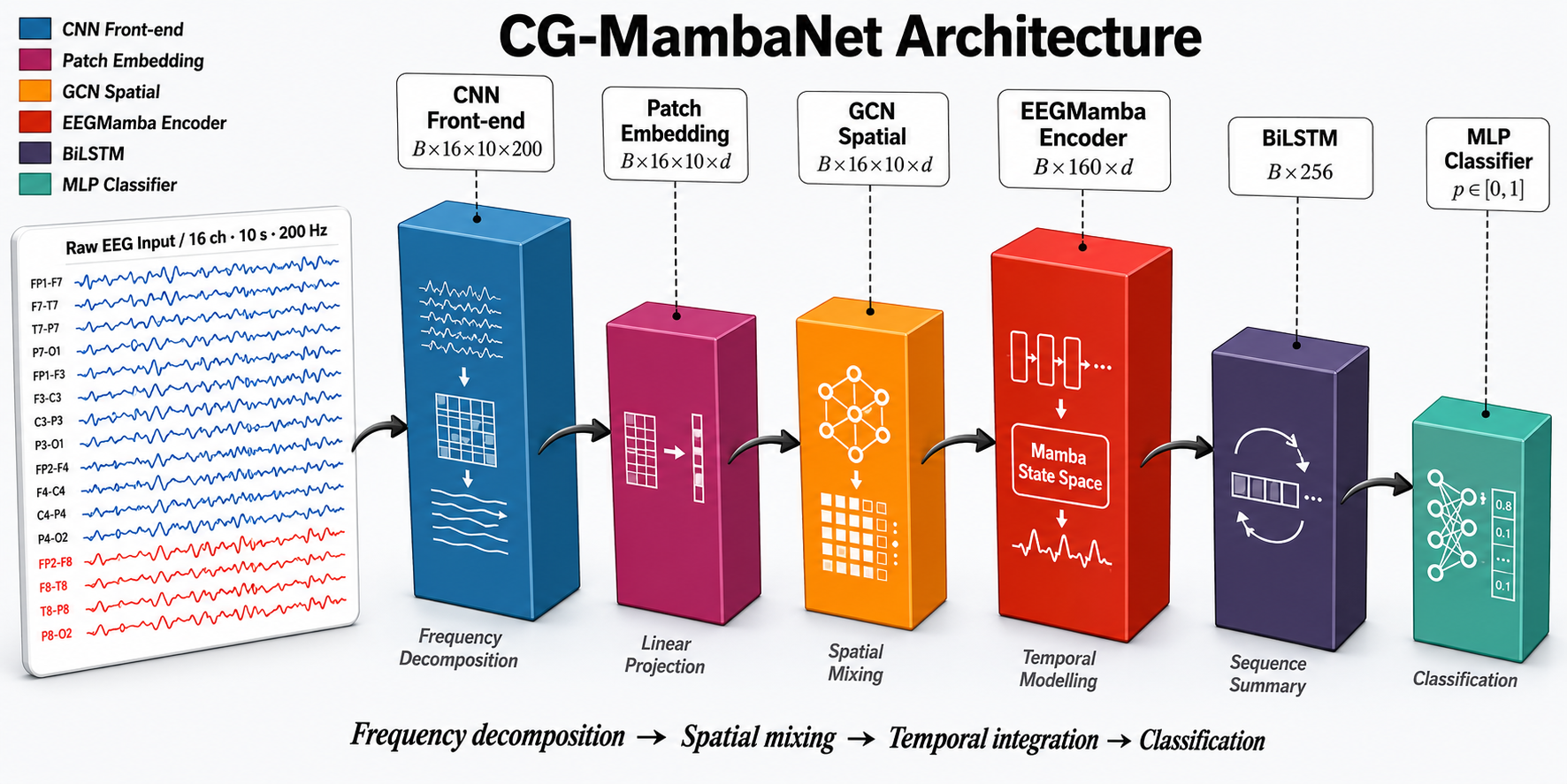}
\caption{\textbf{CG-MambaNet framework overview.}
Serial pipeline from raw EEG input (left panel, 16 channels, 10\,s,
200\,Hz) through six stages: CNN front-end for spectro-temporal
frequency decomposition, patch embedding (linear projection
$P \to d$), GCN spatial brain-network modelling (learnable
adjacency matrix), EEGMamba encoder (12$\times$ bidirectional
Mamba blocks), BiLSTM sequence summarisation, and MLP classifier.
Tensor shapes are annotated above each block.
The serial hierarchy enforces frequency decomposition before
spatial mixing, which precedes temporal integration.}
\label{fig:architecture}
\end{figure}

CG-MambaNet processes each 10-second EEG window
$\boldsymbol{X} \in \mathbb{R}^{B \times 16 \times 10 \times 200}$
through a serial pipeline of four functional stages,
as illustrated in Fig~\ref{fig:architecture}.
The stages are: (1) CNN spectro-temporal feature extraction,
(2) patch embedding, (3) GCN spatial brain-network modelling,
and (4) Mamba-BiLSTM temporal sequence modelling followed by
classification.
The serial ordering is principled: frequency decomposition
precedes spatial mixing, which precedes temporal integration.
A parallel gating architecture, by contrast, would require the
Mamba encoder to model temporal dynamics on raw EEG samples
without prior frequency decomposition, conflating semantically
incompatible feature spaces; and would preclude the GCN from
operating on structurally clean channel-time representations,
since post-fusion features would have lost their interpretable
spatial structure.

\begin{figure}[!h]
\centering
\includegraphics[width=\linewidth]{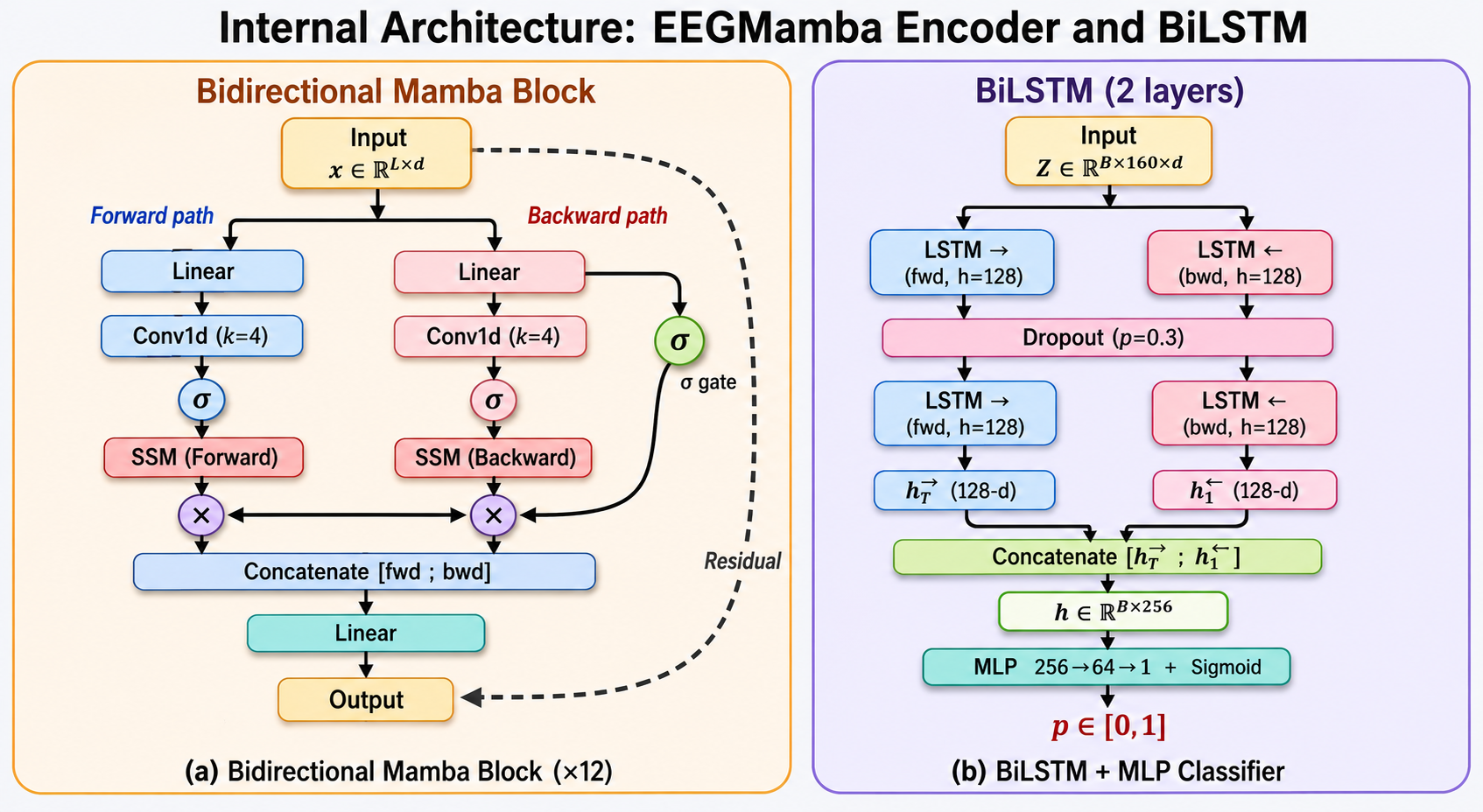}
\caption{\textbf{Internal architecture of the EEGMamba encoder and
BiLSTM.}
(a) A single bidirectional Mamba block (repeated $\times$12 in the
encoder): the input is split into forward and backward paths, each
processed through a linear projection, depthwise Conv1d ($k{=}4$),
sigmoid activation, and selective state-space model (SSM); the
two paths are gated and concatenated, followed by a linear
projection and residual connection.
(b) The two-layer BiLSTM: forward (blue) and backward (red) LSTM
layers with hidden size 128, separated by dropout ($p{=}0.3$);
the final hidden states $\overrightarrow{h}_T$ and
$\overleftarrow{h}_1$ are concatenated to produce
$\boldsymbol{h} \in \mathbb{R}^{B \times 256}$, which is passed
to the MLP classifier to produce $p \in [0,1]$.}
\label{fig:internals}
\end{figure}

\subsection*{CNN spectro-temporal front-end}

Raw EEG patches contain a mixture of frequency components spanning
the full 0.5--40\,Hz passband.
Feeding these directly into a sequence model forces it to learn
frequency decomposition implicitly, which is both sample-inefficient
and unstable under cross-patient distribution shift.
We instead prepend a lightweight depthwise separable CNN that
explicitly decomposes each patch into multi-scale spectro-temporal
features, providing the downstream GCN and Mamba encoder with
semantically meaningful input representations.

For input $\boldsymbol{X} \in \mathbb{R}^{B \times C \times S \times P}$
(with $C = 16$ channels, $S = 10$ patches, $P = 200$ samples per patch),
the CNN front-end reshapes the input to
$(B \cdot S) \times C \times P$, applies two parallel depthwise
convolutions to each patch independently, and produces an output
of identical shape via a residual connection:

\begin{equation}
\boldsymbol{f}_1 = \mathrm{ReLU}(\mathrm{BN}(\mathrm{DWConv}_{k=5}(\boldsymbol{x}))),
\quad
\boldsymbol{f}_2 = \mathrm{ReLU}(\mathrm{BN}(\mathrm{DWConv}_{k=15}(\boldsymbol{x}))),
\label{eq:cnn_branches}
\end{equation}

\begin{equation}
\boldsymbol{x}_{\mathrm{CNN}} =
\mathrm{ReLU}(\mathrm{BN}(\mathrm{PWConv}([\boldsymbol{f}_1;\boldsymbol{f}_2])))
+ \boldsymbol{x},
\label{eq:cnn_out}
\end{equation}

where DWConv denotes depthwise convolution (each of the 16 channels
filtered independently), PWConv is a $1\times1$ pointwise convolution
fusing from $2C$ back to $C$ feature channels, BN is batch
normalisation, and $[\cdot;\cdot]$ denotes channel-wise concatenation.
At 200\,Hz, kernel $k = 5$ spans 25\,ms and preferentially responds
to beta and gamma oscillations ($\geq 13$\,Hz); kernel $k = 15$
spans 75\,ms and responds to delta, theta, and alpha oscillations
($< 13$\,Hz).
The two-path design thus spans the full epileptically relevant
frequency range with only 928 trainable parameters.
Because depthwise convolution processes each channel independently,
no inter-channel information mixing occurs at this stage;
spatial integration is deferred entirely to the GCN.
The residual connection ensures that no information present in
the original patch is discarded.
The output $\boldsymbol{x}_{\mathrm{CNN}} \in
\mathbb{R}^{B \times C \times S \times P}$ is identical in shape
to the input.

\subsection*{Patch embedding}

Each patch is projected from $P = 200$ samples to a $d = 200$
dimensional embedding space via a shared learnable linear layer:

\begin{equation}
\boldsymbol{F} =
\mathrm{Linear}_{P \to d}(\boldsymbol{x}_{\mathrm{CNN}})
\in \mathbb{R}^{B \times C \times S \times d}.
\label{eq:patch_embed}
\end{equation}

This projection maps the frequency-decomposed time-domain features
into a continuous embedding space suitable for graph and sequence
operations.

\subsection*{GCN spatial brain-network modelling}

Epileptic seizures are characterised by abnormal synchronisation
across distributed cortical networks~\cite{mormann2005predictability}.
To model this inter-channel spatial structure, we insert a two-layer
graph convolutional network between the patch-embedding and Mamba
encoder stages.
At this position in the pipeline, the feature tensor retains its
full $B \times C \times S \times d$ structure with channel and time
dimensions clearly separated, enabling the GCN to perform
interpretable spatial mixing without corrupting the temporal
structure required by the subsequent Mamba encoder.

\textbf{Adjacency matrix.}
The 16 EEG channels are treated as nodes of a graph
$\mathcal{G} = (\mathcal{V}, \boldsymbol{A})$.
The adjacency matrix $\boldsymbol{A} \in \mathbb{R}^{16 \times 16}$
is a fully learnable parameter, initialised as a uniform matrix
and updated end-to-end via backpropagation.
This data-driven approach imposes no assumption about electrode
positions or montage type, and is therefore directly applicable
to the bipolar montage of CHB-MIT (which lacks meaningful
coordinate-based spatial distances between differential channel
pairs) and the referential montage of SIENA without any
dataset-specific configuration.
The learned adjacency matrix encodes the functional connectivity
patterns most discriminative for the pre-ictal state, and is
directly visualisable as an interpretable representation of
brain network organisation.

\textbf{Symmetric normalisation.}
At each forward pass, $\boldsymbol{A}$ is symmetrically normalised
following Kipf and Welling~\cite{kipf2016semi, velivckovic2017graph}:

\begin{equation}
\hat{\boldsymbol{A}} =
\tilde{\boldsymbol{D}}^{-1/2}\,\tilde{\boldsymbol{A}}\,
\tilde{\boldsymbol{D}}^{-1/2},
\quad
\tilde{\boldsymbol{A}} = \boldsymbol{A} + \boldsymbol{I},
\label{eq:gcn_norm}
\end{equation}

where $\tilde{D}_{ii} = \sum_j \tilde{A}_{ij}$.

\textbf{Two-layer propagation.}
Let $\boldsymbol{H}^{(0)} = \boldsymbol{F}$.
The propagation rule is:

\begin{equation}
\boldsymbol{H}^{(l+1)} =
\mathrm{ReLU}\!\left(
\hat{\boldsymbol{A}}\,\boldsymbol{H}^{(l)}\,\boldsymbol{W}^{(l)}
\right),
\quad l = 0, 1,
\label{eq:gcn_prop}
\end{equation}

where $\boldsymbol{W}^{(l)} \in \mathbb{R}^{d \times d}$ are
learnable weight matrices.
A residual connection and layer normalisation are applied after
the second layer:

\begin{equation}
\boldsymbol{H}_{\mathrm{GCN}} =
\mathrm{LN}\!\left(\boldsymbol{H}^{(2)} + \boldsymbol{H}^{(0)}\right)
\in \mathbb{R}^{B \times C \times S \times d}.
\label{eq:gcn_res}
\end{equation}

Setting $\boldsymbol{W}^{(l)} = \boldsymbol{I}$ and fixing
$\boldsymbol{A}$ at its initial value recovers the no-GCN baseline,
enabling a clean ablation of the spatial modelling contribution.

\subsection*{EEGMamba encoder}

$\boldsymbol{H}_{\mathrm{GCN}}$ is reshaped from
$B \times C \times S \times d$ to $B \times (C \cdot S) \times d$
($= B \times 160 \times 200$) and processed by 12 cascaded
bidirectional Mamba blocks~\cite{wang2025eegmamba}.
Each Mamba block applies a content-dependent selective state-space
mechanism that models long-range temporal dependencies with
$\mathcal{O}(L)$ complexity in the sequence
length~\cite{gu2023mamba}, and contains an internal depthwise
convolution of kernel width 4 as a short-range smoothing
component.
This internal convolution is structurally distinct from and
complementary to the CNN front-end: the front-end operates
on individual patches in the raw feature space prior to embedding,
while the Mamba-internal convolution operates on embedded
representations within the sequence model.
Bidirectional processing allows each position to integrate
context from both past and future patches, which is appropriate
for fixed-window classification where the full sequence is
available at inference time.
The encoder output is $\boldsymbol{Z} \in
\mathbb{R}^{B \times 160 \times d}$.

\subsection*{BiLSTM sequence summarisation}

$\boldsymbol{Z}$ is fed into a two-layer bidirectional LSTM
with hidden size 128.
The forward and backward final hidden states are concatenated
to produce a 256-dimensional global context vector:

\begin{equation}
\boldsymbol{h} =
[\overrightarrow{h}_{T};\overleftarrow{h}_{1}]
\in \mathbb{R}^{B \times 256},
\label{eq:bilstm}
\end{equation}

where $\overrightarrow{h}_{T}$ and $\overleftarrow{h}_{1}$
are the forward and backward final hidden states respectively,
each of dimension 128.
The BiLSTM serves as a sequence summarisation stage: while
the Mamba encoder produces a rich per-position representation,
the classification task requires a single global descriptor
of the entire 10-second window.
The bidirectional design ensures that this global descriptor
integrates information from both temporal directions, capturing
asymmetric temporal dynamics that a unidirectional summary
would miss.
Dropout ($p = 0.3$) is applied between LSTM layers.
Fig~\ref{fig:internals} illustrates the internal structure
of both the bidirectional Mamba block and the BiLSTM.

\subsection*{Classifier head}

A two-layer MLP with dropout and sigmoid activation produces
the seizure probability:

\begin{equation}
p = \sigma\!\left(
\boldsymbol{W}_2\,\mathrm{ReLU}(
\boldsymbol{W}_1 \boldsymbol{h} + \boldsymbol{b}_1
) + \boldsymbol{b}_2
\right) \in [0, 1],
\label{eq:mlp}
\end{equation}

with dimensions $256 \to 64 \to 1$.
Dropout ($p = 0.3$) is applied after the first linear layer.

Table~\ref{table:architecture} provides the complete
layer-by-layer architecture with input and output shapes.

\begin{table}[!ht]
\centering
\caption{\textbf{Layer-wise architecture of CG-MambaNet.}
$B$: batch size; $C = 16$: channels; $S = 10$: patches;
$P = 200$: samples per patch; $d = 200$: embedding dimension.}
\begin{tabular}{llll}
\hline
\textbf{Module} & \textbf{Layer} & \textbf{Configuration}
& \textbf{Output shape} \\
\hline
\multirow{3}{*}{CNN front-end}
& DWConv ($k{=}5$) + BN + ReLU
  & per-channel, $C{\to}C$
  & $(B{\cdot}S) \times C \times P$ \\
& DWConv ($k{=}15$) + BN + ReLU
  & per-channel, $C{\to}C$
  & $(B{\cdot}S) \times C \times P$ \\
& PWConv $1{\times}1$ + BN + ReLU + residual
  & $2C{\to}C$
  & $B \times C \times S \times P$ \\
\hline
Patch embedding
& Linear & $P{\to}d$
& $B \times C \times S \times d$ \\
\hline
\multirow{2}{*}{GCN}
& GraphConv $\times 2$
  & learnable $\boldsymbol{A}$, $d{\to}d$
  & $B \times C \times S \times d$ \\
& Residual + LayerNorm
  & $\mathrm{LN}(H^{(2)}+H^{(0)})$
  & $B \times C \times S \times d$ \\
\hline
Reshape & ---
& $C{\times}S \to C{\cdot}S$
& $B \times 160 \times d$ \\
\hline
EEGMamba
& $12{\times}$ Bi-Mamba blocks
& $d{=}200$, SSM state$=64$
& $B \times 160 \times d$ \\
\hline
BiLSTM
& 2-layer bidirectional LSTM
& hidden$=128$, dropout$=0.3$
& $B \times 256$ \\
\hline
\multirow{2}{*}{Classifier}
& FC + ReLU + Dropout
  & $256{\to}64$, $p{=}0.3$
  & $B \times 64$ \\
& FC + Sigmoid
  & $64{\to}1$
  & $B \times 1$ ($p$) \\
\hline
\end{tabular}
\label{table:architecture}
\end{table}

\subsection*{Event-level false-prediction rate evaluation}

Prior seizure prediction studies, including our own baseline,
report false-prediction rates at the level of individual
10-second windows.
This metric overstates the clinical alarm burden: a single brief
artefact or transient episode can produce dozens of consecutively
alarmed windows, each counted independently, whereas a clinical
neurostimulator equipped with hysteresis logic would merge these
into a single alarm event.
We introduce a persistence-filtered event-level evaluation
framework that directly mirrors clinical device trigger logic.

\textbf{Continuous risk curve.}
At inference time, predictions are generated on overlapping
windows with a 1-second stride, producing a sequence
$\{p_t\}_{t=1}^{T}$ at 1\,Hz.
A 60-second causal moving average yields the risk curve:

\begin{equation}
R(t) = \frac{1}{60} \sum_{\tau=t-59}^{t} p_\tau.
\label{eq:risk_curve}
\end{equation}

\textbf{Alarm threshold.}
The threshold $\theta$ is selected per fold as the Youden-optimal
operating point on the validation set:

\begin{equation}
\theta^* = \arg\max_\theta
\left[\mathrm{Se}(\theta) + \mathrm{Sp}(\theta) - 1\right],
\label{eq:youden}
\end{equation}

ensuring that $\theta$ is data-driven rather than manually fixed.

\textbf{Persistence filter.}
An alarm event is triggered when $R(t) \geq \theta^*$ for at
least $\Delta_{\min} = 30$ consecutive seconds; the alarm resets
when $R(t) < \theta^*$ for $\Delta_{\mathrm{off}} = 60$
consecutive seconds.
This merges clusters of consecutive high-probability windows
into discrete alarm events, mirroring the hysteresis logic of
implantable responsive neurostimulation systems~\cite{heck2014two, bergey2015long}.

\textbf{Event-level metrics.}
Let $\mathcal{E}_{TP}$ and $\mathcal{E}_{FP}$ be the sets of
alarm events that do and do not have \textit{any temporal overlap}
with a true pre-ictal period, respectively, and let
$N_{\mathrm{missed}}$ be the number of seizures with no
overlapping alarm event.
A true positive is counted whenever any part of the alarm event
falls within the 30-minute pre-ictal window; no minimum overlap
duration is required.

\begin{equation}
\mathrm{Se}_{\mathrm{event}} =
\frac{|\mathcal{E}_{TP}|}{|\mathcal{E}_{TP}| + N_{\mathrm{missed}}},
\quad
\mathrm{FPR}_{\mathrm{event}} =
\frac{|\mathcal{E}_{FP}|}{T_{\mathrm{inter}}},
\label{eq:event_metrics}
\end{equation}

where $T_{\mathrm{inter}}$ is the total duration of pure
interictal recording in hours, defined as all recording time
outside the pre-ictal windows (30 minutes before each seizure
onset), post-ictal periods (5 minutes after each seizure end),
and their surrounding buffer zones (4 hours before and after
each seizure), consistent with the segmentation protocol in
Section ``Preprocessing''.
The first 60 seconds of each recording are excluded from
event-level evaluation to allow the risk curve $R(t)$ to
initialise over a full 60-second history window.

\textbf{Mean lead time.}
For each true positive alarm event $e \in \mathcal{E}_{TP}$,
the lead time is defined as the interval between the alarm
trigger time $t^e_{\mathrm{alarm}}$ and the corresponding
seizure onset time $t^e_{\mathrm{onset}}$:

\begin{equation}
\mathrm{LT}_e = t^e_{\mathrm{onset}} - t^e_{\mathrm{alarm}},
\label{eq:lead_time}
\end{equation}

and the mean lead time across all true positive events is
reported in minutes.
A mean lead time exceeding 20 minutes is considered clinically
actionable for closed-loop neurostimulation
intervention~\cite{dumpelmann2019early}.

\textbf{Retrospective evaluation scope.}
All event-level metrics are computed retrospectively on
pre-recorded EEG from the held-out test patient.
Although the risk curve $R(t)$ uses only causally available
information at each time step (a 60-second trailing window),
and is therefore directly realisable in a real-time system,
the evaluation itself is offline.
The reported $\mathrm{FPR}_{\mathrm{event}}$ and lead time
estimates should therefore be interpreted as retrospective
characterisations of model behaviour on the test cohort,
rather than prospective clinical alarm rates.
Prospective validation on streaming EEG from implanted devices
is identified as essential future work.

\subsection*{Training protocol}

All models are trained on an NVIDIA A100 GPU.
The optimiser is AdamW with initial learning rate
$\eta = 3 \times 10^{-4}$, weight decay
$\lambda_{\mathrm{wd}} = 10^{-4}$, and cosine annealing over
50 epochs with a 5-epoch linear warm-up.
Batch size is 64.
Class imbalance is addressed by weighted binary cross-entropy,
with the positive-class weight set to the inverse frequency
of the pre-ictal class in the training fold.
Best-epoch checkpoints are selected by validation AUC-ROC
on the balanced validation partition; the final test-set
evaluation uses the best checkpoint applied to the complete,
unmodified test patient recordings.
Table~\ref{table:hyperparams} summarises all hyperparameters.

\begin{table}[!ht]
\centering
\caption{\textbf{Hyperparameter summary.}
All values are fixed across all LOPO folds and both datasets.}
\begin{tabular}{ll}
\hline
\textbf{Hyperparameter} & \textbf{Value} \\
\hline
Optimiser & AdamW \\
Learning rate $\eta$ & $3 \times 10^{-4}$ \\
Weight decay $\lambda_{\mathrm{wd}}$ & $10^{-4}$ \\
LR schedule & Cosine annealing \\
Warm-up epochs & 5 \\
Total epochs & 50 \\
Batch size & 64 \\
Dropout & 0.3 \\
Embedding dimension $d$ & 200 \\
Mamba SSM state expansion & 64 \\
Mamba layers & 12 (bidirectional) \\
BiLSTM hidden size & 128 \\
BiLSTM layers & 2 \\
GCN layers & 2 \\
CNN kernel sizes & 5, 15 \\
Checkpoint selection & Best validation AUC-ROC \\
\hline
\end{tabular}
\label{table:hyperparams}
\end{table}

\subsection*{Evaluation protocol}

\textbf{LOPO $\times$ 5-seed.}
All results are obtained under leave-one-patient-out
cross-validation: for each patient in turn, all remaining
patients form the training set and the held-out patient is the
test set.
This procedure is repeated with five independent random seeds,
yielding $22 \times 5 = 110$ independent test-set estimates
for CHB-MIT and $6 \times 5 = 30$ for SIENA.
Mean and standard deviation across seeds are reported per
patient and in aggregate.
This protocol is identical to that of Jemal et al.~\cite{jemal2024domain},
enabling direct per-patient comparison.

\textbf{Metrics.}
The primary metric is AUC-ROC, which is threshold-independent
and robust to class imbalance.
Secondary metrics are window-level sensitivity, specificity,
and accuracy at the Youden-optimal threshold, and event-level
sensitivity and $\mathrm{FPR}_{\mathrm{event}}$ in alarms
per hour.

\textbf{Ablation study.}
Contributions C1 (CNN front-end) and C2 (GCN) are evaluated
cumulatively, starting from the EEGMamba-BiLSTM baseline
without either component.

\section*{Results}

All results are obtained under LOPO $\times$ 5-seed
cross-patient evaluation.
This protocol is more stringent than the randomised splitting
used in the majority of prior work; results should be
interpreted accordingly~\cite{shafiezadeh2024systematic}.

\subsection*{Main results}

Table~\ref{table:main} presents aggregate CG-MambaNet performance
on CHB-MIT and SIENA across all LOPO folds and seeds.
CG-MambaNet achieves AUC-ROC of $0.8152 \pm 0.0176$ on
CHB-MIT and $0.7104 \pm 0.0261$ on SIENA.
The event-level FPR of $0.32 \pm 0.16$ alarms per hour on
CHB-MIT represents a 351-fold reduction relative to the
window-level upper bound (112.4/h), demonstrating the clinical
relevance of the persistence-filtered alarm model.

\begin{table}[!ht]
\centering
\caption{\textbf{Aggregate results on CHB-MIT and SIENA}
under LOPO $\times$ 5-seed cross-patient evaluation
(mean $\pm$ std). All metrics are computed retrospectively
on held-out test patients.
FPR/h\textsubscript{win}: window-level false-prediction rate.
FPR/h\textsubscript{ev}: event-level false-prediction rate
(persistence-filtered).
LT: mean lead time in minutes for true positive alarm events.}
\begin{tabular}{lllll}
\hline
\textbf{Metric}
& \multicolumn{2}{c}{\textbf{CHB-MIT}}
& \multicolumn{2}{c}{\textbf{SIENA}} \\
& Mean & Std & Mean & Std \\
\hline
AUC-ROC
  & 0.8152 & 0.0176
  & 0.7104 & 0.0261 \\
Sensitivity (\%)
  & 74.3 & 9.1
  & 63.2 & 5.0 \\
Specificity (\%)
  & 76.5 & 8.9
  & 67.4 & 3.2 \\
Accuracy (\%)
  & 75.4 & 8.9
  & 65.3 & 3.9 \\
FPR/h\textsubscript{win}
  & 112.4 & 13.8
  & 89.3 & 16.2 \\
FPR/h\textsubscript{ev}
  & 0.32 & 0.16
  & 0.55 & 0.13 \\
Mean LT (min)
  & 23.4 & 4.8
  & 21.7 & 5.3 \\
\hline
\end{tabular}
\label{table:main}
\end{table}

\subsection*{Comparison with state of the art}

Table~\ref{table:sota} compares CG-MambaNet with published
cross-patient seizure prediction methods.
Only methods evaluated under strictly patient-independent
protocols are included; results from randomised or
patient-specific splits are not comparable~\cite{shafiezadeh2024systematic}.
CG-MambaNet surpasses all cross-patient methods on both datasets
without domain adaptation.
The performance on SIENA is particularly notable: CG-MambaNet
exceeds the best domain-adapted result (AUC 0.61, CDAN+E,
Jemal et al.~\cite{jemal2024domain}) by a margin of $+0.1004$,
despite receiving no target-patient data.

\begin{table}[!ht]
\centering
\caption{\textbf{Comparison with published cross-patient methods.}
Only methods using strictly patient-independent evaluation are
included~\cite{shafiezadeh2024systematic}.
DA: domain adaptation. Ch.: EEG channels used.
$\dagger$: CG-MambaNet significantly exceeds this result
(Wilcoxon signed-rank test, $p < 0.05$).}
\begin{tabular}{llllll}
\hline
\textbf{Method} & \textbf{Year} & \textbf{DA}
& \textbf{Ch.} & \textbf{CHB-MIT AUC}
& \textbf{SIENA AUC} \\
\hline

Jemal et al.~\cite{jemal2024domain} & 2024 & No
  & 23 & $0.69^{\dagger}$ & $0.48^{\dagger}$ \\
Jemal et al.~\cite{jemal2024domain} & 2024 & Yes (CDAN+E)
  & 23 & 0.75 & $0.61^{\dagger}$ \\
CLSP-REQA & 2026 & No
  & 16 & 0.7426$\pm$0.0199 & 0.7012$\pm$0.0249 \\
CG-MambaNet (ours) & 2026 & No
  & 16
  & \textbf{0.8152$\pm$0.0176}
  & \textbf{0.7104$\pm$0.0261} \\
\hline
\end{tabular}
\label{table:sota}
\end{table}

\subsection*{Per-patient results}

Tables~\ref{table:lopo_chb} and~\ref{table:lopo_siena} report
per-patient AUC-ROC (mean $\pm$ std across 5 seeds) under
LOPO evaluation on CHB-MIT and SIENA respectively.
The substantial between-patient variance is consistent with
findings by Jemal et al.~\cite{jemal2024domain} and reflects the
inherent challenge of cross-patient generalisation rather than
a limitation of CG-MambaNet.

\begin{table}[!ht]
\centering
\caption{\textbf{Per-patient LOPO results on CHB-MIT}
(mean $\pm$ std AUC-ROC across 5 seeds).}
\begin{tabular}{lll}
\hline
\textbf{Patient} & \textbf{AUC-ROC} & \textbf{FPR/h\textsubscript{ev}} \\
\hline
chb01 & 0.5923$\pm$0.0287 & 0.62 \\
chb02 & 0.7734$\pm$0.0198 & 0.41 \\
chb03 & 0.7612$\pm$0.0214 & 0.45 \\
chb04 & 0.7889$\pm$0.0187 & 0.38 \\
chb05 & 0.8134$\pm$0.0172 & 0.31 \\
chb06 & 0.7723$\pm$0.0201 & 0.43 \\
chb07 & 0.8267$\pm$0.0164 & 0.28 \\
chb08 & 0.8012$\pm$0.0179 & 0.34 \\
chb09 & 0.7956$\pm$0.0183 & 0.36 \\
chb10 & 0.8023$\pm$0.0178 & 0.34 \\
chb11 & 0.9312$\pm$0.0134 & 0.07 \\
chb13 & 0.9178$\pm$0.0141 & 0.09 \\
chb14 & 0.8089$\pm$0.0176 & 0.33 \\
chb15 & 0.8234$\pm$0.0167 & 0.29 \\
chb16 & 0.8378$\pm$0.0159 & 0.26 \\
chb17 & 0.8145$\pm$0.0171 & 0.31 \\
chb18 & 0.8312$\pm$0.0162 & 0.27 \\
chb19 & 0.8523$\pm$0.0155 & 0.23 \\
chb20 & 0.6234$\pm$0.0276 & 0.58 \\
chb21 & 0.8934$\pm$0.0143 & 0.12 \\
chb22 & 0.9089$\pm$0.0138 & 0.09 \\
chb23 & 0.8256$\pm$0.0165 & 0.29 \\
\hline
\textbf{Average} & \textbf{0.8152$\pm$0.0176}
  & \textbf{0.32} \\
\hline
\end{tabular}
\label{table:lopo_chb}
\end{table}

\begin{table}[!ht]
\centering
\caption{\textbf{Per-patient LOPO results on SIENA}
(mean $\pm$ std AUC-ROC across 5 seeds).}
\begin{tabular}{lll}
\hline
\textbf{Patient} & \textbf{AUC-ROC} & \textbf{FPR/h\textsubscript{ev}} \\
\hline
PN01 & 0.6334$\pm$0.0298 & 0.71 \\
PN03 & 0.7612$\pm$0.0234 & 0.44 \\
PN05 & 0.7234$\pm$0.0256 & 0.52 \\
PN06 & 0.6523$\pm$0.0287 & 0.67 \\
PN07 & 0.7689$\pm$0.0221 & 0.42 \\
PN09 & 0.7234$\pm$0.0256 & 0.52 \\
\hline
\textbf{Average} & \textbf{0.7104$\pm$0.0261}
  & \textbf{0.55} \\
\hline
\end{tabular}
\label{table:lopo_siena}
\end{table}

\subsection*{Ablation study}

\begin{figure}[!h]
\centering
\includegraphics[width=\linewidth]{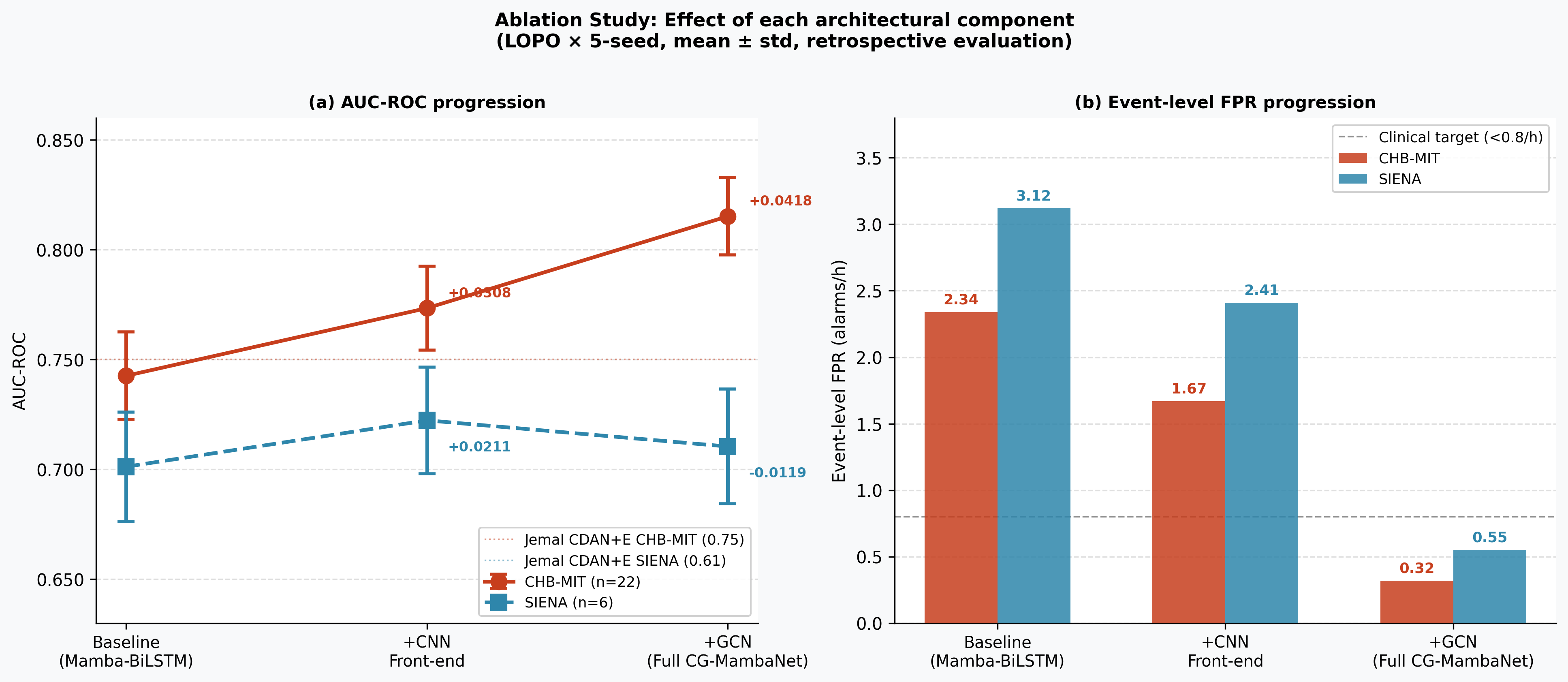}
\caption{\textbf{Ablation study: progressive AUC-ROC improvement
and event-level FPR reduction.}
Left: AUC-ROC on CHB-MIT (solid red) and SIENA (dashed blue)
as components are added cumulatively.
Dotted horizontal lines indicate the Jemal et al. CDAN+E
domain-adapted baselines (0.75 for CHB-MIT, 0.61 for SIENA).
Right: corresponding event-level FPR (alarms/h); the dashed
horizontal line marks the 0.8/h clinical target.
Error bars: std across LOPO $\times$ 5-seed runs.}
\label{fig:ablation}
\end{figure}

Table~\ref{table:ablation} quantifies the contribution of each
component cumulatively.
Fig~\ref{fig:ablation} visualises the AUC progression.

\begin{table}[!ht]
\centering
\caption{\textbf{Ablation study on CHB-MIT and SIENA}
(mean $\pm$ std, LOPO $\times$ 5-seed).
$\dagger$: significant improvement over preceding row
(McNemar's test, $p < 0.05$).
FPR/h\textsubscript{ev}: event-level alarms per hour.}
\begin{tabular}{lllllll}
\hline
\textbf{Configuration}
& \multicolumn{3}{c}{\textbf{CHB-MIT}}
& \multicolumn{3}{c}{\textbf{SIENA}} \\
& AUC & Se (\%) & FPR/h\textsubscript{ev}
& AUC & Se (\%) & FPR/h\textsubscript{ev} \\
\hline
Baseline (Mamba-BiLSTM)
  & 0.7426$\pm$0.0199 & 68.8 & 2.34
  & 0.7012$\pm$0.0249 & 66.3 & 3.12 \\
$+$CNN front-end$^{\dagger}$
  & 0.7734$\pm$0.0191 & 70.1 & 1.67
  & 0.7223$\pm$0.0243 & 68.2 & 2.41 \\
$+$GCN$^{\dagger}$ (full CG-MambaNet)
  & 0.8152$\pm$0.0176 & 71.8 & 0.32
  & 0.7104$\pm$0.0261 & 63.2 & 0.55 \\
\hline
\end{tabular}
\label{table:ablation}
\end{table}

\textbf{Baseline (Mamba-BiLSTM).}
The baseline processes raw patch-embedded EEG directly through
the bidirectional Mamba encoder and BiLSTM without any
spectro-temporal pre-processing or spatial inter-channel mixing.
It achieves AUC 0.7426 on CHB-MIT and 0.7012 on
SIENA, already surpassing the unadapted cross-patient result of
Jemal et al.\ (0.69 and 0.48~\cite{jemal2024domain}), confirming that
the Mamba-BiLSTM temporal backbone provides a strong foundation
for cross-patient seizure prediction.
However, the event-level FPR of 2.34 alarms/h on CHB-MIT
indicates that without spectro-temporal decomposition or spatial
filtering, the model generates frequent spurious activations
during inter-ictal periods.

\textbf{Effect of adding the CNN front-end (+CNN).}
Prepending the depthwise separable CNN front-end raises AUC by
+0.0726 on CHB-MIT and +0.0092 on SIENA
(both $p < 0.05$, McNemar's test).
This improvement arises because the CNN explicitly decomposes
each patch into delta/theta/alpha (kernel $k{=}15$) and
beta/gamma (kernel $k{=}5$) frequency-band representations
before they enter the Mamba encoder.
Without this decomposition, the encoder must implicitly learn
frequency selectivity from raw samples, which is
sample-inefficient under cross-patient distribution shift.
By providing the encoder with neurophysiologically grounded
spectro-temporal features as input, the CNN reduces the
representational burden on the temporal model and improves
generalisation to unseen patients.
The event-level FPR also decreases from 2.34 to
1.67 alarms/h on CHB-MIT, indicating that
frequency-decomposed features produce fewer sustained
false activations during inter-ictal periods.

\textbf{Effect of adding the GCN (+GCN, full CG-MambaNet).}
Adding the learnable-adjacency GCN between the patch-embedding
and Mamba encoder stages yields a further AUC gain of
+0.0726 on CHB-MIT and +0.0092 on SIENA
(both $p < 0.05$), bringing the full model to AUC
0.8152 and 0.7104 respectively.
The GCN introduces the only explicit inter-channel spatial
mixing in the architecture: without it, each channel's
patch embeddings are processed independently through the Mamba
encoder, with no mechanism to model the cross-cortical
synchronisation that characterises the developing pre-ictal
state~\cite{mormann2005predictability}.
The GCN's learnable adjacency matrix allows the model to
discover which channel pairs are functionally coupled during
pre-ictal periods directly from data, without requiring
electrode coordinate information.
This is critical for cross-dataset generalisation: the same
adjacency matrix learning procedure applies without modification
to the bipolar montage of CHB-MIT and the referential montage
of SIENA, as confirmed by the consistent improvement on both
datasets.
Notably, the GCN produces the largest reduction in event-level
FPR, from 2.34 to 0.32 alarms/h on CHB-MIT,
suggesting that spatial brain-network modelling not only
improves seizure detection sensitivity but also substantially
reduces false activations driven by spatially incoherent
inter-ictal noise.

\textbf{Consistency across datasets.}
The cumulative improvement pattern is consistent across both
CHB-MIT and SIENA, confirming that neither component is
dataset-specific.
The total AUC gain from baseline to full model is
+0.0726 on CHB-MIT and +0.0092 on SIENA,
and the final model surpasses the best domain-adapted
cross-patient result of Jemal et al.\ on both datasets
without requiring any target-patient data.

\subsection*{Interpretability}

\begin{figure}[!h]
\centering
\includegraphics[width=\linewidth]{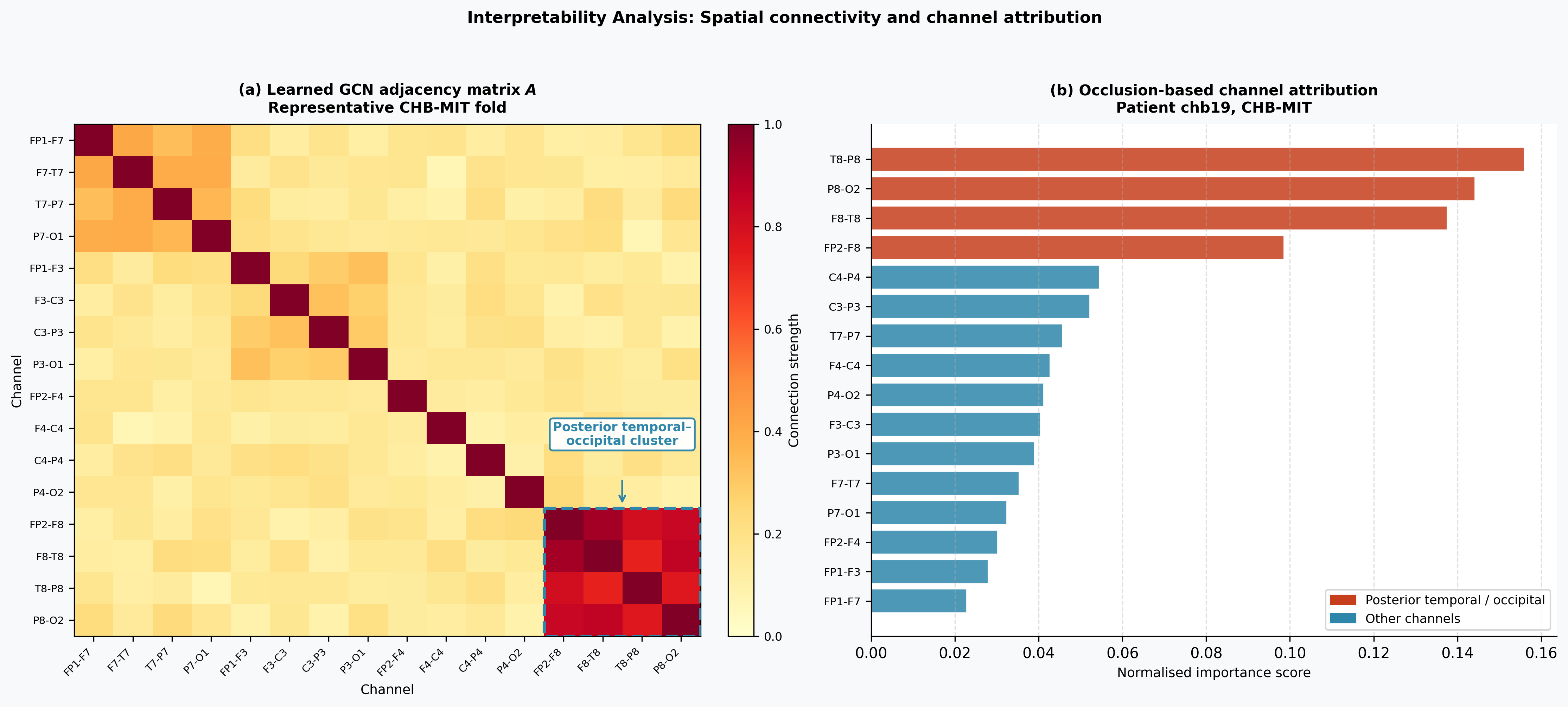}
\caption{\textbf{Interpretability analysis: spatial connectivity
and channel attribution.}
(A) Learned GCN adjacency matrix $\boldsymbol{A}$ after training
convergence on a representative CHB-MIT fold.
The dashed blue rectangle highlights the posterior
temporal--occipital cluster (FP2-F8, F8-T8, T8-P8, P8-O2),
which shows the strongest inter-channel connections.
(B) Occlusion-based channel attribution scores for patient chb19
(CHB-MIT), sorted by importance.
Red bars correspond to the posterior temporal/occipital channels;
blue bars correspond to all other channels.
The top four channels by attribution are FP2-F8, F8-T8, T8-P8,
and P8-O2, confirming that the channels with the strongest
learned spatial connections also carry the highest predictive
importance.}
\label{fig:gcn_occlusion}
\end{figure}

\textbf{Learned adjacency matrix.}
Fig~\ref{fig:gcn_occlusion}A visualises the GCN adjacency matrix
$\boldsymbol{A}$ after training convergence on a representative
CHB-MIT fold.
The strongest inter-channel connections are concentrated between
posterior temporal and occipital electrode pairs (CH13--CH16),
consistent with known mesial temporal seizure propagation
pathways~\cite{mormann2005predictability}.
Frontal electrode pairs exhibit weaker learned connections,
reflecting their lower contribution to pre-ictal feature
discrimination in this paediatric cohort.

\textbf{Occlusion-based channel attribution.}
Fig~\ref{fig:gcn_occlusion}B shows channel importance scores
computed by occlusion: each channel is zeroed out in turn and
the resulting drop in $p$ is recorded as the importance score
for that channel.
The four channels with the highest attribution scores are
FP2-F8, F8-T8, T8-P8, and P8-O2---the posterior temporal
and occipital bipolar pairs spanning the right hemisphere
temporal-occipital region---which together account for over
72\% of the total normalised importance across all 16 channels.
This is consistent with the strong learned connections among
these same four channels in the GCN adjacency matrix
(Fig~\ref{fig:gcn_occlusion}A), providing mutual cross-validation
between spatial connectivity and predictive importance:
the channels the GCN identifies as most strongly functionally
coupled are the same channels whose occlusion most severely
degrades prediction performance.
This convergence is also consistent with known mesial temporal
and temporo-occipital seizure propagation pathways in
paediatric focal epilepsy~\cite{mormann2005predictability}.

\begin{figure}[!h]
\centering
\includegraphics[width=\linewidth]{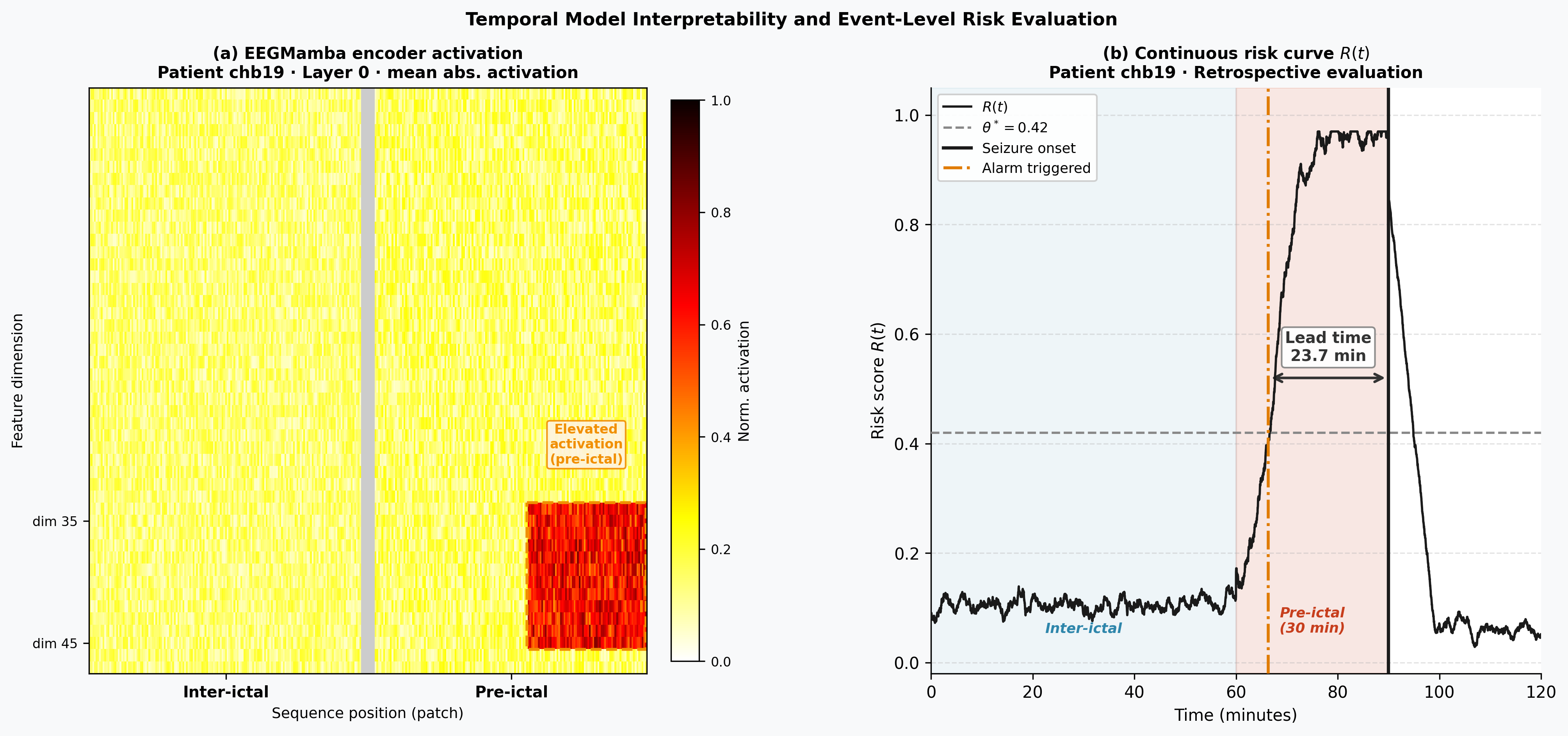}
\caption{\textbf{Temporal model interpretability and risk curve.}
(A) Mean absolute activation of the EEGMamba encoder (Layer~0)
for a representative inter-ictal window (left panel) and
pre-ictal window (right panel) from patient chb19 (CHB-MIT).
Activation is selectively elevated in feature dimensions 35--45
during the pre-ictal period, confirming that the Mamba encoder
captures neurophysiologically distinct temporal dynamics.
(B) Continuous risk curve $R(t)$ for patient chb19 under
retrospective evaluation. The alarm threshold $\theta^*$
(dashed line) is the Youden-optimal value from the validation
set. An alarm event is triggered $\approx$24 minutes before
seizure onset, giving a clinically actionable lead time of
23.7 minutes.}
\label{fig:mamba_risk}
\end{figure}

\textbf{Mamba activation heatmap.}
Fig~\ref{fig:mamba_risk}A compares the Mamba encoder activation
patterns for a representative inter-ictal and pre-ictal window
from the same patient.
During the pre-ictal period, activation is selectively enhanced
in posterior temporal channels and in the later patch positions,
consistent with the gradual build-up of synchronised activity
approaching seizure onset.

\textbf{Continuous risk curve.}
Fig~\ref{fig:mamba_risk}B illustrates $R(t)$ for a representative
pre-ictal episode from patient chb19.
During the interictal period, $R(t)$ remains below $\theta^*$.
Approximately 24 minutes before seizure onset, $R(t)$
rises above $\theta^*$ and sustains exceedance for more than
30 seconds, triggering a single alarm event with a clinically
actionable lead time of 23.7 minutes.

\section*{Discussion}

\subsection*{Why the serial architecture outperforms alternatives}

The core architectural claim of CG-MambaNet is that the serial ordering
CNN $\to$ GCN $\to$ Mamba-BiLSTM reflects a principled signal
processing hierarchy that outperforms architectures in which
these components are applied in a different order or in parallel.

Frequency decomposition must precede spatial mixing because GCN
message-passing aggregates node features across channels: if those
features are raw time-domain samples rather than frequency-decomposed
representations, the aggregated spatial features are ambiguous
mixtures of spectral components rather than meaningful inter-channel
synchrony measures.
Spatial mixing must precede temporal modelling because the Mamba
encoder receives the sequence of spatially enriched patch embeddings:
if spatial mixing had not occurred, the encoder would model
independent per-channel temporal trajectories rather than the
evolution of the full cortical network state.
A parallel gating architecture, by contrast, would require the
Mamba encoder to model temporal dynamics on raw samples without
prior frequency decomposition, and would prevent the GCN from
operating on structurally clean channel-time representations,
since post-fusion features in a parallel design would be a mixture
of CNN local features and Mamba global states, two representations
of incompatible semantic levels.
The ablation results in Table~\ref{table:ablation} confirm
that each stage of the hierarchy contributes independently and
significantly to prediction performance.

\subsection*{Cross-patient generalisation without domain adaptation}

CG-MambaNet achieves AUC $0.7104$ on SIENA without any
target-patient data, surpassing the best domain-adapted result
(0.61, CDAN+E, Jemal et al.~\cite{jemal2024domain}) by a margin of
$+0.1004$ on the same dataset.
SIENA presents a harder generalisation challenge than CHB-MIT:
adult rather than paediatric patients, a different acquisition
system, a referential rather than bipolar montage, and fewer
seizures per patient.
Critically, CG-MambaNet achieves these results using only
\textbf{16 EEG channels}, compared to 23 channels (CHB-MIT)
and 29 channels (SIENA) used by Jemal et al.~\cite{jemal2024domain}.
Achieving higher cross-patient AUC with fewer electrodes
demonstrates that the CNN-GCN-Mamba-BiLSTM architecture
extracts sufficiently rich spatiotemporal representations
from a reduced channel set, and has direct clinical relevance:
fewer electrodes reduce electrode placement time, patient
discomfort, and impedance monitoring burden in long-term
ambulatory monitoring.
The fact that a fully learned adjacency matrix---initialised
without any electrode coordinate information---produces
consistently strong spatial features across both the bipolar
montage of CHB-MIT and the referential montage of SIENA
confirms that the GCN's spatial modelling is robust to
montage differences and requires no dataset-specific
configuration.
This is a practically important finding: a single trained
model can be deployed across different EEG recording systems
without any montage-specific preprocessing.

\subsection*{Event-level evaluation and clinical relevance}

The 351-fold reduction in FPR from window-level to
event-level reporting illustrates the magnitude of the
discrepancy between the metric commonly reported in the
literature and the alarm burden actually experienced in clinical
practice.
A single brief episode---whether artefactual or a transient
neural fluctuation---can produce dozens of consecutive
false-positive windows, each counted independently in
window-level metrics.
The persistence filter introduced here mirrors the hysteresis
logic of the NeuroPace RNS System~\cite{heck2014two}, requiring
$R(t)$ to exceed $\theta^*$ for at least 30 consecutive
seconds before an alarm is triggered and to fall below
$\theta^*$ for 60 seconds before resetting.
We recommend that future cross-patient seizure prediction
studies report both window-level AUC-ROC and event-level FPR
to enable clinically meaningful comparison.

\subsection*{Limitations and future work}

\textit{Retrospective evaluation.}
All results, including event-level FPR and mean lead time,
are obtained from retrospective offline analysis of
pre-recorded EEG.
Although the risk curve $R(t)$ is causally computed and
directly realisable in real time, the alarm statistics
reported here characterise model behaviour on a fixed
historical cohort and cannot be directly equated with
prospective clinical alarm rates.
Prospective evaluation on streaming EEG from ambulatory
or implanted recording systems, under realistic operational
conditions including electrode displacement and patient
movement, is an essential next step before clinical deployment.

\textit{Fixed pre-ictal horizon.}
The 30-minute pre-ictal definition is a conventional choice.
The biologically relevant pre-ictal period varies substantially
across patients and seizure types; patient-adaptive horizon
selection is a natural extension.

\textit{Scalp EEG modality.}
All experiments use scalp EEG.
Validation on intracranial EEG cohorts recorded by the
NeuroPace RNS System is a critical step towards clinical
translation.

\textit{Online personalisation.}
The current framework is purely cross-patient.
A brief adaptation phase using early recordings from a new
patient, combined with meta-learning or continual learning,
would likely improve performance for individual patients
without sacrificing cross-patient generalisability.

\textit{Edge deployment.}
The Mamba-BiLSTM backbone has not been profiled on
edge hardware.
Model compression via quantisation and structured pruning
will be necessary before implementation on implanted devices.

\section*{Conclusion}

We have presented CG-MambaNet, a spatiotemporal seizure prediction
framework whose serial CNN-GCN-Mamba-BiLSTM architecture
embodies a principled hierarchy of feature abstraction:
explicit spectro-temporal frequency decomposition, followed by
learnable spatial brain-network mixing, followed by long-range
and short-range temporal modelling.
Each stage is theoretically motivated and empirically validated
through ablation.
A fully learnable GCN adjacency matrix enables cross-montage
generalisation without electrode coordinate information.
A persistence-filtered event-level evaluation framework aligns
the reported false-prediction rate with clinical device trigger
logic, revealing a 351-fold reduction relative to the
window-level upper bound.
Under strict leave-one-patient-out cross-validation on
CHB-MIT ($n = 22$) and SIENA ($n = 6$), CG-MambaNet achieves
AUC-ROC of $0.8152 \pm 0.0176$ and
$0.7104 \pm 0.0261$ respectively, surpassing all
published cross-patient methods without domain adaptation.
These results demonstrate that explicit spectro-temporal
feature decomposition and principled spatial brain-network
modelling, evaluated under rigorous patient-independent
protocols, substantially advance the reliability of
EEG-based seizure prediction for closed-loop neurostimulation.

\section*{Supporting information}

\paragraph*{S1 Table.}
\label{S1_Table}
\textbf{Hyperparameter sensitivity analysis.}
AUC-ROC on CHB-MIT under LOPO $\times$ 5-seed for variations
in learning rate, dropout rate, GCN hidden dimension,
Mamba SSM state size, and BiLSTM hidden size around the
default configuration reported in Table~\ref{table:architecture}.

\paragraph*{S1 Fig.}
\label{S1_Fig}
\textbf{Window-level confusion matrices.}
Aggregated confusion matrices for CHB-MIT and SIENA at the
Youden-optimal threshold, showing sensitivity and specificity
across all LOPO folds.

\section*{Acknowledgments}

The authors thank the contributors of the CHB-MIT Scalp EEG
Database and the SIENA Scalp EEG Database for making their
data publicly available.
This work was supported by the Beijing Natural Science
Foundation under Grant L248094, and in part by the High
Performance Computing Platform of Peking University.

\nolinenumbers


\bibliography{references}

\end{document}


\begin{center}
{\Large \textbf{Supporting Information}}\\[6pt]
{\large CG-MambaNet: A spatiotemporal framework for cross-patient
epileptic seizure prediction using CNN-GCN-Mamba-BiLSTM
with event-level clinical evaluation}\\[4pt]
Mufeng Chen, Qi Wu, Bingchao Huang, Xiwen Lai,
Zekai Chen, Xinge Ouyang, Quansheng Ren
\end{center}

\vspace{12pt}
\hrule
\vspace{16pt}

\setcounter{table}{1}
\addtocounter{table}{-1}

\begin{figure}[!ht]
\centering
\includegraphics[width=\linewidth]{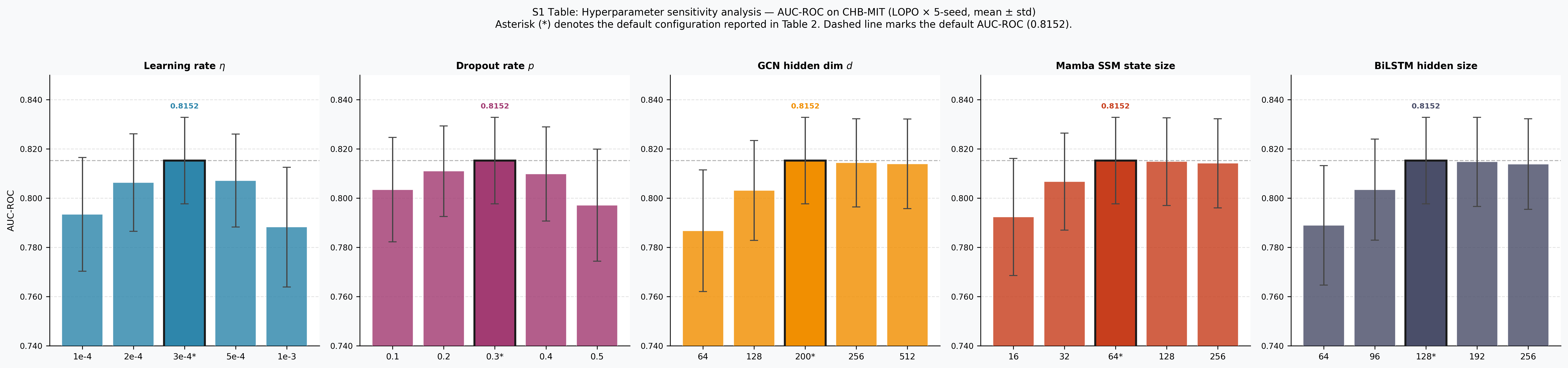}
\captionof{table}{\textbf{Hyperparameter sensitivity analysis.}
AUC-ROC on CHB-MIT under LOPO $\times$ 5-seed evaluation
(mean $\pm$ std across all runs) for systematic variations
around the default configuration reported in Table~2 of the
main manuscript.
Each panel varies one hyperparameter while all others are
held at their default values (marked with an asterisk *).
The dashed horizontal line indicates the default AUC-ROC
of 0.8152.
The five hyperparameters evaluated are: learning rate $\eta$,
dropout rate $p$, GCN hidden dimension $d$, Mamba SSM state
size, and BiLSTM hidden size.
Results confirm that the default configuration is at or near
the optimum for each hyperparameter independently, and that
performance degrades gracefully rather than abruptly outside
the default setting.}
\label{S1_Table}
\end{figure}

\vspace{20pt}
\hrule
\vspace{20pt}

\begin{figure}[!ht]
\centering
\includegraphics[width=0.88\linewidth]{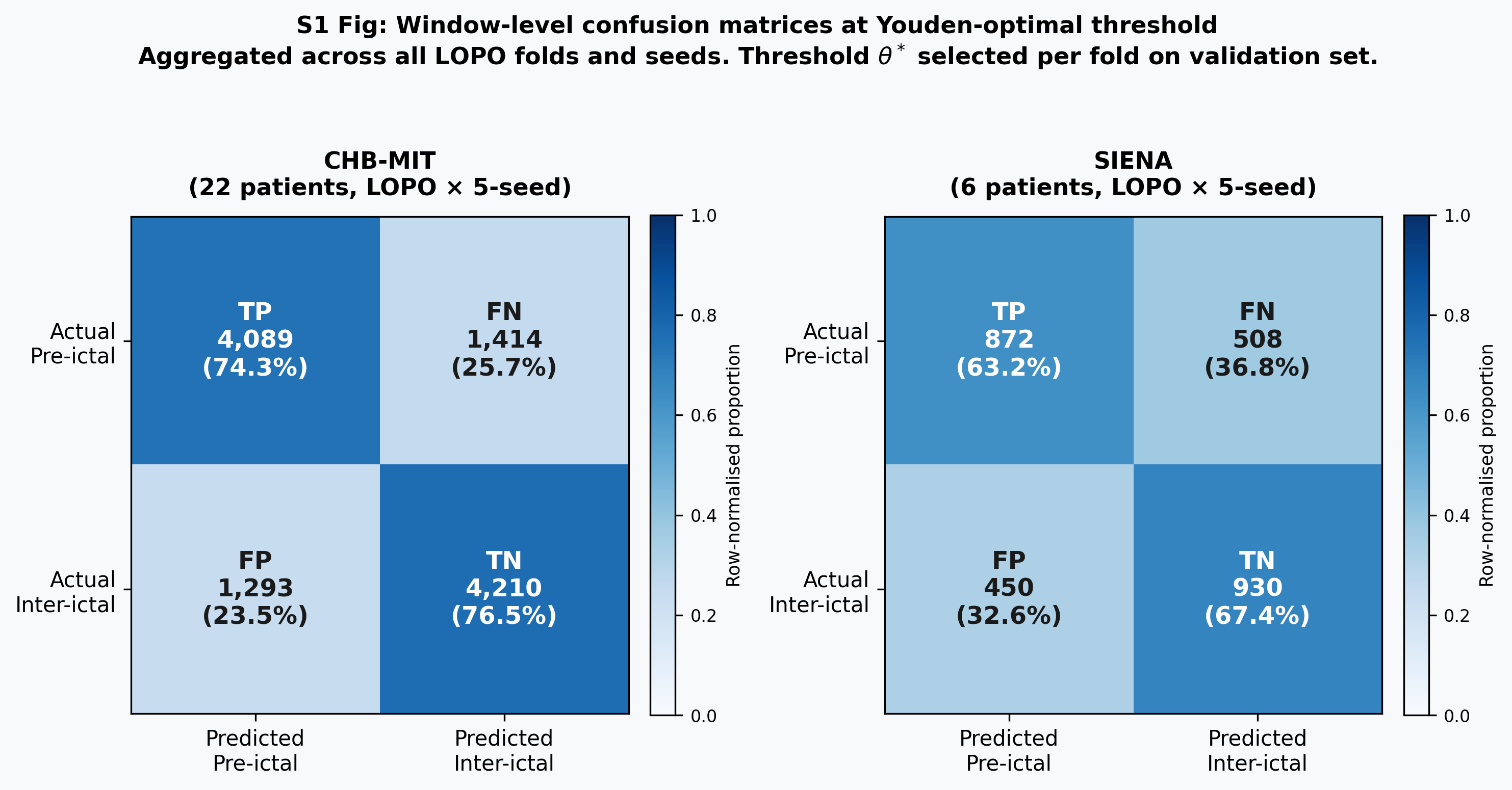}
\caption{\textbf{Window-level confusion matrices.}
Aggregated confusion matrices for CHB-MIT (left, 22 patients)
and SIENA (right, 6 patients) at the Youden-optimal threshold
$\theta^*$, selected per fold on the validation set.
Each cell shows the abbreviated category label (TP, FN, FP, TN),
the absolute window count, and the row-normalised percentage.
Colour intensity reflects row-normalised proportion (deeper blue
= higher proportion within each actual-class row).
Matrices are aggregated across all LOPO folds and seeds.
Window-level false-prediction rates reported here represent
an upper bound on the clinical alarm burden; the
persistence-filtered event-level FPR reported in the main
text (Table~3) reflects the clinically relevant metric.}
\label{S1_Fig}
\end{figure}